\documentclass{PoS}
\usepackage{amsmath}
\usepackage{braket}
\usepackage{float}

\def\be{\begin{equation}}
\def\ee{\end{equation}}
\def\bea{\begin{eqnarray}}
\def\eea{\end{eqnarray}}
\title{How $\mathcal N=1$, $D=4$ SYM domain walls look like}

\ShortTitle{SYM domain walls}

\author{Igor Bandos$^{a,b}$, Stefano Lanza$^{c}$
and \speaker{Dmitri Sorokin}$^{d}$\\
\llap{$^a$}Department of
	Theoretical Physics, University of the Basque Country UPV/EHU\\
P.O. Box 644, 48080 Bilbao\\
\llap{$^b$}IKERBASQUE, Basque Foundation for Science\\
48011 Bilbao, Spain\\
\llap{$^c$}Jefferson Physical Laboratory, Harvard University\\
Cambridge, MA 02138, USA\\
\llap{$^d$}INFN, Padova Section\\
via F. Marzolo 8, 35131 Padova, Italy\\
E-mail: \email{igor.bandos@ehu.eus}, \email{slanza@g.harvard.edu},
\email{sorokin@pd.infn.it}}

\abstract{We review main features of the pure $\mathcal N=1$, $D=4$ SYM and its effective description by the Veneziano-Yankielowicz generalized sigma-model. We then indicate that the construction of  BPS domain
walls interpolating between different SYM vacua requires the presence of a dynamical membrane
source. We will show how such a membrane is coupled to the SYM and present the explicit form of
BPS domain walls which it creates in the Veneziano-Yankielowicz effective theory. In particular, we will describe 1/2 BPS domain wall configurations with $|k|\leq N/3$, where $k$ is the membrane charge that sets the ``distance'' between two distinct SUSY vacua.
}

\FullConference{Corfu Summer Institute 2019 "School and Workshops on Elementary Particle Physics and Gravity" (CORFU2019)\\
		31 August - 25 September 2019\\
		Corfu, Greece}

\begin{document}

\section{Introduction}
The $\mathcal N=1$, $D=4$ supersymmetric Yang-Mills was constructed in 1974 \cite{Wess:1974jb,Ferrara:1974pu,Salam:1974ig}
and studied intensively over 45 years. In spite of its seeming simplicity, this theory has revealed a rich quantum structure. For instance,
since early `80s it has been known that e.g. pure $\mathcal N=1$, $D=4$ SYM with a gauge group $SU(N)$ has $N$ degenerate susy vacua distinguished by different vacuum expectation values of the gluino condensate \cite{Witten:1982df,Shifman:1987ia,Davies:1999uw}
which are related to each other by discrete R-symmetry transformations,
\begin{equation}\label{gc}
\braket{{\rm Tr}\, \lambda^\alpha\lambda_\alpha}= \Lambda^3 e^{2\pi i \frac nN}, \quad n=0,1, \ldots N-1,
\end{equation}
where $\rm Tr$ denotes the trace with respect to gauge symmetry indices and $\Lambda$ is a dynamical scale at which the condensate of the gluini $\lambda_\alpha$ is formed by non-perturbative effects.

On general grounds, it was then suggested that there should exist BPS domain walls interpolating between different SYM vacua (labeled by $l$ and $n$) and having the following tension saturating the BPS bound \cite{Dvali:1996xe}
\be\label{dwT}
T_{\rm DW}=\frac N{8\pi^2}\,|\braket{\lambda\lambda}_{n}-\braket{\lambda\lambda}_{l}|\,.
\ee
Since the `90s, domain walls in $\mathcal N=1$, $D=4$ SYM and Supersymmetric QCD have been extensively studied with the use of different approaches (for a recent review and latest developments see e.g. \cite{Bashmakov:2018ghn}). However, explicit solitonic solutions of a low-energy effective field theory describing domain walls in the pure SYM  have not been found until recently. A reason for this is that SYM  BPS domain walls are not smooth solitonic field configurations. Their existence requires the presence of a source which has been lacking in pure ${\mathcal N=1}$, $D=4$ SYM. In \cite{Kogan:1997dt} it was suggested that such an object (associated with integrated heavy modes of the theory) should contribute to the BPS value \eqref{dwT} of the domain wall tension. In \cite{Bandos:2019qok} we have shown that this object is a dynamical membrane. The purpose of this contribution is to show how to couple the membrane to ${\mathcal N=1}$ SYM and its Veneziano-Yankielowicz effective field theory formulation \cite{Veneziano:1982ah}, how the membrane creates BPS domain walls and what is their shape.

\section{Pure ${\mathcal N=1}$ $SU(N)$ SYM }

To set up the stage and point out essential blocks of our construction, let us briefly review the main features of the simple SYM theory with the gauge group $SU(N)$. Its field content is a vector supermultiplet in the adjoint representation of $SU(N)$
\be\label{vm}
A^I_m,\,\, \lambda^I_\alpha,\,\, \bar\lambda^I_{\dot\alpha},\,\, D^I\; ,
\ee
where $A^I_m(x)$ $(m=0,1,2,3)$ is the vector gauge field, $\lambda^I_\alpha(x)$ $(\alpha =1,2)$and its complex conjugate $\bar\lambda^I_{\dot\alpha}$ $({\dot\alpha}=1,2)$ are the Weyl spinor gluino,
% and
$D^I(x)$ is a space-time scalar auxiliary field
and $I=1,...,N^2-1$ is the index of the adjoint representation of $SU(N)$.
In the rest of the paper we will skip these $SU(N)$ indices over the fields, i.e.  we will consider $su(N)$ algebra valued fields $A_m(x)=A^I_m(x){\tt T}_I$, etc.  where ${\tt {T}}_I$ are the $su(N)$ generators.

The building block of the SYM action is chiral spinor superfield which in the chiral superspace basis, parametrized by complex Grassmann-odd coordinates $\theta_\alpha$ and  $x_L^m=x^m+i\theta\sigma^m\bar{\theta}$,
has the following form
\be\label{calW}
\mathcal W_\alpha(x_L,\theta)=-i\lambda_\alpha +\theta_\alpha D- \frac i 2 F_{mn}\sigma^{mn}{}_\alpha{}^\beta\theta_\beta
+\theta^2\sigma^{m}_{\alpha\dot\beta}
\nabla_m\bar{\lambda}{}^{\dot\beta}, %\quad \bar D_{\dot\alpha}{\mathcal W}_\beta=0,
\ee
where $F_{mn}$ is the gauge field strength, $\nabla_m=\partial_m-iA_m$ is the gauge covariant derivative and $\sigma^{m}_{\alpha\dot\beta}$ are the relativistic Pauli matrices.

The $\mathcal N=1$ SYM Lagrangian is an integral of the square of $\mathcal W_\alpha$ over $\theta$
\be\label{SYML}
\mathcal L_{\rm SYM}=\frac 1{4g^2}\, \int d^2\theta \,{\rm Tr}\, {\mathcal W}^\alpha {\mathcal W}_\alpha \,+\, {\rm c.c.},
\ee
where $g$ is the SYM coupling constant.

Note that the Lagrangian is invariant under the $U(1)$ R-symmetry ${\mathcal W}_\alpha \to e^{i\varphi} {\mathcal W}_\alpha$, which is broken by quantum chiral anomalies to a discrete subgroup $\mathbb Z_{2N}$. Namely, the R-symmetry current conservation becomes anomalous
\be\label{RA}
\partial_mJ^m:=\partial_m\,{\rm Tr}\,(\lambda\sigma^m\bar\lambda)=\frac {2N}{32\pi^2} \varepsilon^{mnpq}\,{\rm Tr}\,F_{mn}F_{pq}.
\ee
 The factor of $2N$ appears on the right hand side of the above equation since $\lambda$ are in the adjoint of $SU(N)$. As a result the generating functional of the quantum theory is invariant under a residual $\mathbb Z_{2N}$ symmetry. The latter is however further broken down to $\mathbb Z_2$ ($\lambda \to -\lambda$) due to the formation of the gluino condensate \eqref{gc} by non-perturbative effects.

\subsection{SYM Lagrangian and the special chiral superfield}
For the possibility of coupling the membrane to the SYM multiplet it is important to notice that ${\rm Tr}\, {\mathcal W}^\alpha {\mathcal W}_\alpha$ is a chiral scalar superfield (which is special as we will see in a minute)
\be\label{S=WW}
S={\rm Tr}\, {\mathcal W}^\alpha {\mathcal W}_\alpha =s+\sqrt 2\theta^\alpha\chi_\alpha+\theta^2F,
\ee
where
\bea\label{s}
s&=&-{\rm Tr}\, \lambda^\alpha \lambda_\alpha\,,
\\[10pt]
\label{chi}
\chi_\alpha&=&\sqrt{2} {\rm Tr}\, \left(\frac 1 2 \, F_{mn}\sigma^{mn}_{\alpha}{}^\beta\lambda_\beta-i\,\lambda_\alpha D \right),
\eea
and
\be\label{F=2}
F={\rm Tr}\,\left(-2i\lambda\sigma^m \nabla_m\bar\lambda-\frac 12F_{mn}F^{mn}+D^2 -\frac i 4 \varepsilon_{mnpl}F^{mn}F^{pl}\right)\,.
\ee
Note that the real part of the F-term is the SYM Lagrangian, while its imaginary part is an instanton density which is (locally) a differential of a Chern-Simons three-form, namely
\bea\label{F4dCS}
F_4=d^4x\,\,{{\rm Im\hskip0.1em}} F&=&-{\rm Tr}\, F_2\wedge F_2-d^4x\, \partial_m ({\rm Tr}\, \lambda\sigma^m\bar\lambda)  \qquad \\ &=&- d{\rm Tr} \left(A d A+\frac {2i}3 A^3+\frac 1 {3!}
 d x^k d x^nd x^m \epsilon_{mnkl}{\rm Tr} \,\lambda\sigma^l\bar\lambda \right)\equiv d C_3.\nonumber
\eea
{(For shortness, we have omitted the wedge product symbols in the second line).}
Therefore, the complex field $F$ in \eqref{S=WW} has the following form
\be\label{F}
F=\hat D+i\,*dC_3=\hat D+i\,\partial_mC^m,
\ee
where $\hat D$ is a scalar field and $C^m$ is the Hodge dual of the three-form $C_3$ ($C_1=*C_3$). Hence, the F-term is gauge invariant under $C_3 \to C_3+d\Lambda_2$ with $\Lambda_2(x)$ being a two-form gauge parameter.

This structure of the F-term makes the chiral superfield \eqref{S=WW} special, as was noticed e.g. in \cite{Burgess:1995kp,Binetruy:1996xw}. Superfields of this kind were first considered by Gates in \cite{Gates:1980ay}. They can always be expressed as a second super-covariant derivative of a real scalar `prepotential' $U(x,\theta,\bar\theta)$
\be\label{S=D2U}
S=-\frac 14 \bar D_{\dot\alpha}\bar D^{\dot\alpha} U.
\ee
In the case of the conventional generic chiral superfields the prepotential $U$ is complex.

The superfield $U$ contains the components of the real one-form  $C_1=dx^mC_m$ dual to $C_3$ among its independent bosonic components
\be\label{U}
\begin{aligned}
U |_{\theta=\bar\theta=0} &= u ,\\
- \frac 18 \bar{\sigma}^{\dot\alpha \alpha}_m [D_\alpha, \bar{D}_{\dot\alpha}] U|_{\theta=\bar\theta=0} &= C_m,\\
\frac1{4} D^2 U|_{\theta=\bar\theta=0} &= -\bar s={\rm Tr} \,\bar\lambda\bar\lambda,\\
\frac{1}{16} D^2 \bar{D}^2 U|_{\theta=\bar\theta=0} &= \hat D + i \partial^m C_m \equiv F \, .
\end{aligned}
\ee
Note that the superfield $S$ in \eqref{S=D2U} is invariant under the superfield transformation
\be\label{UtoU+L}
U'= U+L\,,
\ee
where $L$ is a real linear superfield, i.e. the superfield satisfying
\be
\label{Lcon}
\bar D^2L=0=D^2L\,.
\ee
This transformation is the superfield extension of the gauge variation of the three-form $C_3\to C_3 + d\Lambda_2$ under which the F-tern in \eqref{U} is invariant. Hence, only gauge-invariant combinations of the components of $U$ appear in $S$.

As we will see later, the presence of the three-form $C_3$ in $U$ allows one to couple it to a membrane. Since we will look for domain walls of SYM induced by the membranes in its effective Veneziano-Yankielowicz field theory description, let us now revisit the structure of the latter.

\section{Veneziano-Yankielowicz Lagrangian and potential}
The VY Lagrangian \cite{Veneziano:1982ah} provides an effective description of colorless bound states of the SYM multiplet (like glueballs and gluinoballs), and demonstrates the formation of the gluino condensate and the N-degeneracy of the $SU(N)$ SYM vacuum. The form of the VY Lagrangian is (almost) fixed by anomalous superconformal Ward identities of the SYM. Its building block is the colorless chiral superfield \eqref{S=WW} whose components \eqref{s}-\eqref{F}  are now regarded as independent fields, rather than composites of the SYM multiplet. The VY Lagrangian has the following form
\be\label{VYL}
\mathcal L_{\rm VY}= \frac 1{16\pi^2 \rho} \int d^2\theta d^2\bar \theta (S\bar S)^{\frac 13}+\int d^2\theta\, W(S)+{\rm c.c.}\,.
\ee
The first term in \eqref{VYL} contains the K\"ahler potential
\be\label{K=}
K(S, \bar S)= \frac 1{16\pi^2 \rho} (S\bar S)^{\frac 13}
\ee
with a priori arbitrary dimensionless positive constant $\rho$. Its simplest form is chosen due to the mass dimension 3 of the superfield $S$.

In general, the kinetic part of the Lagrangian is not fixed by anomalous symmetries and can also include higher order terms, however only for the above choice of the K\"ahler potential the scalar field potential is bounded from below \cite{Shore:1982kh}.

The second term in \eqref{VYL} is the VY superpotential. It is uniquely fixed by  anomalous superconformal Ward  identities of the SYM theory and has the following form
\be\label{VYsp}
W(S)=\frac N{16\pi^2}S\left(\ln \frac{S}{\Lambda^{3}}-1\right)\,, \qquad W_S:= \partial_SW(S)=\frac N{16\pi^2}\ln \frac{S}{\Lambda^{3}}\,.
\ee
However, the superpotential and hence the Lagrangian are not single valued under an identical phase transformation of $S$. Indeed, they shift as
\be\label{StoS'}
S \to S\,e^{2\pi\,i}\,, \qquad {\mathcal L}_{VY} \to {\mathcal L}_{VY}-\frac N{4\pi}\,{\partial_m C^m}.
\ee
Another (related) issue is that the F-term in $S$ is not a complex auxiliary scalar field but contains the dual four-form field strength $\partial_m C^m$ associated with the SYM instanton density ${\rm Tr}\,F_2\wedge F_2$. So the integration of the auxiliary field $F$ out of the VY action requires caution.
A recipe of how one can take care of these subtleties by modifying the VY superpotential was proposed in \cite{Kovner:1997im}.
Instead, we will follow  a somewhat different procedure of augmenting the VY Lagrangian \cite{Bandos:2019qok}  prompted by general requirements of the consistent construction of $4D$ Lagrangians containing three-form gauge fields (see e.g. \cite{Groh:2012tf,Farakos:2017jme} and \cite{Lanza:2019nfa} for a recent review). Namely, the special form of the chiral superfield  $S$ \eqref{S=D2U} requires the variation of the VY Lagrangian with respect to  the independent real superfield $U$. The variation principle is well-defined only with the addition of the boundary (total derivative) term which for the case under consideration is \footnote{A general prescription for constructing such terms was given in \cite{Farakos:2017jme}.}
\be\label{bdtS}
{\mathcal L}_{\rm bd}
=-\frac 1{128\pi^2} \left(\int\,d^2\theta \bar D^2-\int\,d^2\bar\theta D^2\right)\left[\left(\frac 1{12\rho } \bar D^2 \frac{\bar S^\frac 13}{{S^\frac 23}}{+}\ln \frac{\Lambda^{3N}}{S^N}\right)U\right]
+\text{c.c.}
\ee
It is not hard to see that the Lagrangian
\be\label{VY+bd}
\mathcal L=\mathcal L_{\rm VY}+\mathcal L_{\rm bd}
\ee
is invariant not only under the identical phase transformation \eqref{StoS'} but under a continuous $U(1)$ symmetry.  To break this symmetry down to $Z_{2N}$, as it occurs in the SYM, we will require that the term $X(S{{, \bar S}})\equiv \frac 1{16\pi^2 }\left(\frac 1{12\rho } \bar D^2 \frac{\bar S^\frac 13}{{S^\frac 23}}{+} \ln \frac{\Lambda^{3N}}{S^N}\right)$ in the Lagrangian (\ref{bdtS})  satisfies the following boundary conditions
\be\label{Xbc}
X(S{{, \bar S}})|_{\rm bd}={-}\frac{i\, n}{8\pi}, \qquad {\rm where}\qquad  n=0,1\ldots, (N-1)\,({\rm{mod}}\, N)
 \ee
which characterizes the asymptotic vacua of the theory. Note that with this choice of the boundary conditions the Lagrangian (\ref{VY+bd}) is gauge invariant under the superfield transformation \eqref{UtoU+L}.

We are interested in studying classical configurations of fields in the VY model with no fermionic excitations. Thus we set $\chi_\alpha=0$ in \eqref{VYL} and \eqref{bdtS}, and get the following bosonic Lagrangian
\be\label{bosonicVYL}
{\mathcal L}^{\rm bos}_{\rm VY}=  K_{s\bar s}\left({- \partial_m{s}\partial^m{\bar s}}+(\partial_m C^{m})^2+\hat D^2 \right)
 + \left( {W}_s \left(\hat D+i\partial_m C^{m}  \right) + \text{c.c.}\right) + {\mathcal L}^{\rm bos}_{\rm bd}\,
\ee
where the boundary term has the following form
\be\label{bosoncibt}
{\mathcal L}^{\rm bos}_{\rm bd} =  - 2\partial_m \left[ C^{m} \left(  K_{s\bar s}  \partial_n C^{n}
- {\rm Im}\, W_{s}\right) \right]\,
\ee
and
$$
 K_{s\bar s}\equiv \partial_s\partial_{\bar s}K(s,\bar s), \quad W_s\equiv \partial_s W(s).
$$
We will now eliminate the fields $\hat D$ and $C^m$ from the Lagrangian by solving their equations of motion. For the field $\hat D$ we have
\be\label{hatD}
K_{s\bar s} \hat D+{\rm Re}\, W_s=0\qquad \to \qquad \hat D=-\frac {{\rm Re}\,W_s}{K_{s\bar s}}
\ee
and for $C^m$
\be\label{Cm}
\partial_m(K_{s\bar s} \partial_nC^n-{\rm Im}\, W_s)=0 \qquad \to \qquad \partial_mC^m=\frac {{\rm Im}\, W_s-{\frac{n}{8\pi}}}{K_{s\bar s}}\,.
\ee
So
\be\label{Fexpress}
F\equiv\hat D+{\rm i} \partial_mC^m=-\frac {{\overline W}_{\bar s}+{{\rm i}\,\frac{n}{8\pi}}}{K_{s\bar s}}=-\frac {\partial_{\bar s}({\overline W}+{{\rm i}\,\frac{n}{8\pi}}\bar s)}{K_{s\bar s}},
\ee
where $\frac n{8\pi}$ with $(n=0,1,\ldots, N-1) $ (mod $N$) is the integration constant compatible with the choice \eqref{Xbc} of the boundary conditions.

The right hand side of \eqref{Cm} implies that the original VY superpotential gets effectively shifted by a term linear in $S$, i.e.
\be\label{Wjump}
W(S) \to W(S)-{{\rm i}\,\frac{n}{8\pi}}S\,.
\ee
  Substituting the expression \eqref{Fexpress} for $F$ into the action \eqref{bosonicVYL} we get the scalar field potential first derived in \cite{Kovner:1997im}
\be\label{sV}
V(s,\bar s)
  =\frac{9 \rho N}{16\pi^2}\,|s|^{\frac 43}\left(\ln^2\frac {|s|}{\Lambda^{3}}+(\arg s - {2\pi \frac nN})^2\right)\,, \quad n =0,  1,  2, \ldots, N-1 \,\, ({\rm mod}\,\,N)\,.
\ee
In this potential the parameter $n$ should be considered as a discrete variable. This makes the potential single-valued and multi-branched (periodic in $n$ with the period $N$). It has cusps at $\arg s=\frac {\pi (k+1)}N$, at which $n$ changes its value from $k$ to $k+1$, and becomes zero (namely, it has absolute SUSY minima) at
$\langle s\rangle=\Lambda^3 e^{2\pi {\rm i}\frac nN}$. The latter effectively reproduce the gluino condensate \eqref{gc} of the $SU(N)$ SYM.

\begin{figure}[h!t]
    \centering
	\includegraphics[width=10cm]{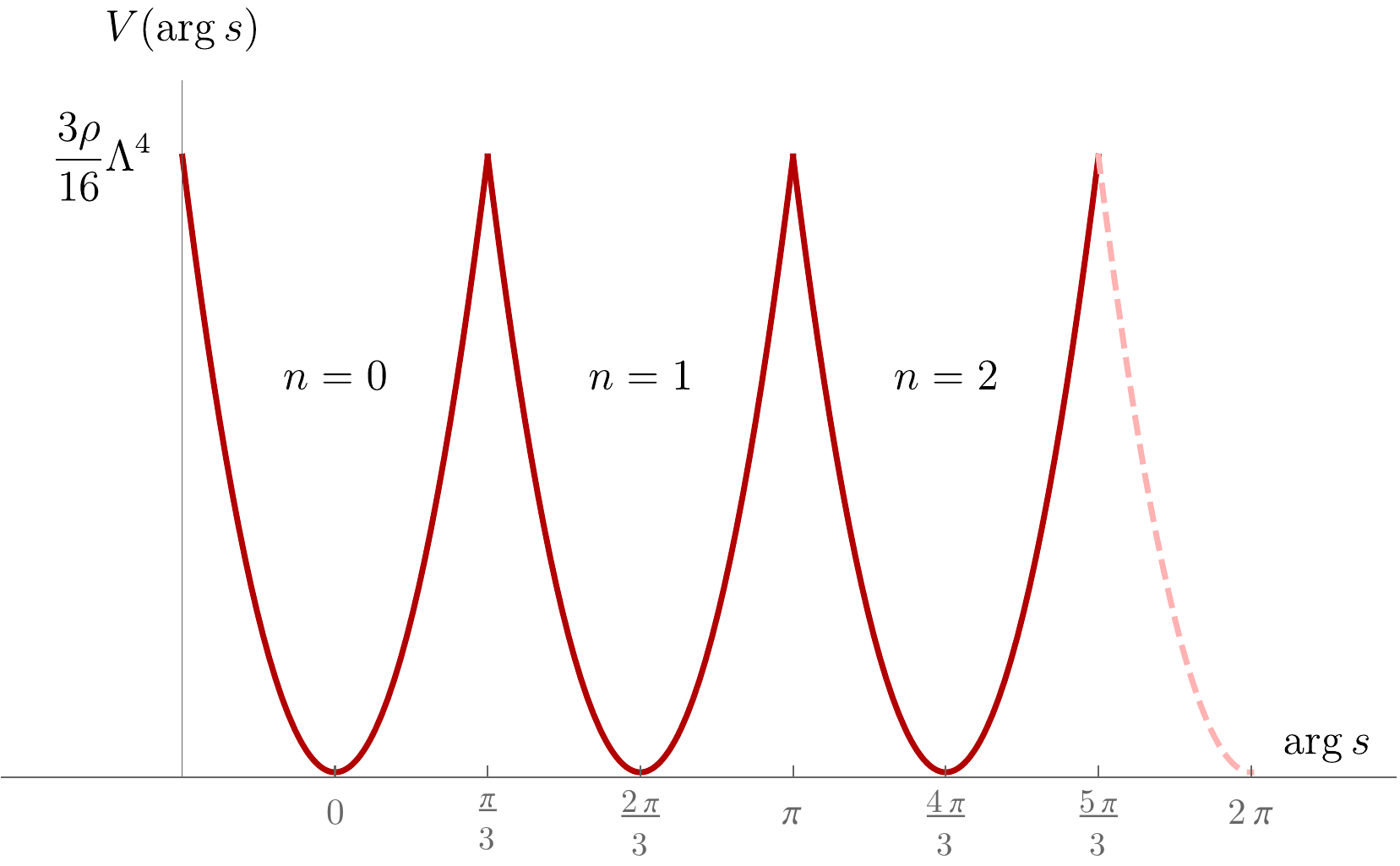}
	
	\caption{\footnotesize
The VY potential as a function of $\arg s$ for $|s|=\Lambda^3$ and $N=3$. The dashed line to the right should be identified with that to the very left. The potential is zero (has the absolute minima) at $\arg{s}=\frac{2\pi n}{3}$ $(n=0,1,2)$ and cusps at $\arg{s}=\frac{\pi (n+1)}{3}$  at which the variable $n$ changes its value. 	}
	
	\label{fig:Potential}
\end{figure}

For instance, for $|s|=\Lambda^3$ and $N=3$ for which $\frac \pi N \simeq 1$, the potential has the dependence on $\arg s$ and $n$ that is depicted on Figure~\ref{fig:Potential}.

The presence of cusps indicates that at the corresponding points of space there sits an object that causes $n$ to change its value and correspondingly the superpotential to ``jump'' as in \eqref{Wjump}.  We will now show that this object is a dynamical membrane which carries a quantized charge that couples it to the three-form gauge field.

\section{Coupling the membrane to the SYM and VY model}

Membranes in flat and supergravity superspaces of various dimensions, and their effects in string/M-theory have been studied at length since the 80's after a seminal paper \cite{Bergshoeff:1987cm}. It was soon realized \cite{Townsend:1987yy,Abraham:1990nz} that in $D=4$ supersymmetric membranes have to do with BPS-saturated domain walls. In this respect it is somewhat surprising that the problem of coupling dynamical membranes to $\mathcal N=1$, $D=4$ SYM has been addressed only recently in \cite{Bandos:2019qok}, though back in the early `90s generic constructions of couplings of $p$-branes to bosonic Young-Mills fields in various dimension were proposed in \cite{Dixon:1991xz,Dixon:1992qd} and a bosonic membrane coupled to $4D$ gauge fields via the Chern-Simons term (which is closely related to our supersymmetric construction) was considered in \cite{Townsend:1993wy}.

The action describing the dynamics of a membrane coupled to the $\mathcal N=1$ SYM or its VY effective field model is a generalization of the supermembrane action of \cite{Bandos:2010yy} coupled to a special chiral superfield \eqref{S=D2U}. It has the following form
\be\label{susym1}
S_{\rm membrane}=-\frac 1{4\pi}\int_{{\mathcal M}_3}d^3\,\xi\sqrt{-\det h_{ij}}\left|kS+c\right| -\frac k{4\pi}\int_{{\mathcal M}_3}{\mathcal C}_3-\left(\frac{\bar c}{4\pi}\int_{{\mathcal M}_3}{\mathcal C}^0_3+c.c.\right),
\ee
where  $c=k_1+{\rm i} k_2$, and $k$, $k_1$ and $k_2$ are real constant charges characterizing the membrane coupling to a real three-form gauge superfield ${\mathcal C}_3$ and a complex super three-form ${\mathcal C}^0_3$ to be defined below. The  normalization factor $\frac  1{4\pi}$ has been chosen to have the canonical form of the Chern-Simons term in the static membrane action which forces the charge $k$  be quantized $k=\pm 1,\pm 2\,\ldots$.

In the Nambu-Goto part of the action \eqref{susym1}  the bulk superfield $S(x_L,\theta)$ is either a composite special chiral superfield \eqref{S=WW} or its Veneziano-Yankielowicz counterpart.\footnote{The coupling of a membrane to an Abelian gauge field supermultiplet (associated with R-symmetry) within $\mathcal N=1$, $D=4$ supergravity, which enters a special chiral superfield in a different way has been recently considered in \cite{Cribiori:2020wch}.} It is evaluated on the membrane worldvolume $z^M= z^M(\xi)$ parametrized by $\xi^i\,(i=0,1,2)$.
The constant $c$ added to $S$ insures that when $S={\rm Tr}\,W^\alpha W_\alpha$ ({which is} a nilpotent superfield, $S^{2N}\equiv 0$) the module $|S+c|$ is well defined. In the VY model in which $S$ is a fully-fledged special chiral superfield, we will for simplicity set $c=0$.

The induced metric on the membrane worldvolume is
\be
h_{ij}(\xi)=%\textcolor{green}{:=}
\eta_{ab}E^a_i(\xi)E^b_j(\xi),\quad~~\text{with }\quad E^a_i(\xi){:=} \partial_i z^M(\xi)E^a_M(z(\xi)),
\ee
and
\be\label{sva}
E^a(\xi)=d z^M(\xi)E_M^a(z(\xi)){=: d\xi^i E^a_i(\xi)}= d x^a(\xi)+i\theta\sigma^a d\bar\theta(\xi)-i\, d\theta\sigma^a \bar\theta(\xi)
\ee
is the worldvolume pull-back of the flat superspace vector supervielbein.

The super three-form $\mathcal C_3$ is constructed in terms of the real prepotential $U$ (see eqs. \eqref{S=D2U} and \eqref{U})
\be\label{super3form}
\begin{aligned}
\mathcal C_{3}=&\,   {i} E^a \wedge d\theta^\alpha \wedge d\bar\theta^{\dot\alpha}  \sigma_{a\alpha\dot\alpha}U \\ & - {\frac 14}  E^b\wedge E^a \wedge  d\theta^\alpha
\sigma_{ab\; \alpha}{}^{\beta}{D}_{\beta}U -{\frac 14}  E^b\wedge E^a \wedge  d \bar\theta^{\dot\alpha}
\bar\sigma_{ab}{}^{\dot\beta}{}_{\dot\alpha}\bar{D}_{\dot\beta}U
\\&-\frac {1} {48}
  E^c \wedge E^b \wedge E^a \epsilon_{abcd} \,\bar{\sigma}{}^{d\dot{\alpha}\alpha}
  [D_\alpha, \bar{D}_{\dot\alpha}]U
 \, .
\end{aligned}
\ee
Note that the last, purely tensorial, term in \eqref{super3form} coincides, at $\theta=\bar\theta=0$, with the three-form dual of the vector component of $U$ in \eqref{U}. Specifically, in the SYM case this term is nothing but the Chern-Simons term \eqref{F4dCS}. So the leading bosonic component of the Wess-Zumino term $\mathcal C_3$ is
$$
{\mathcal C}_3|_{\theta=0}={C_3}=-{\rm Tr} \left(AdA+\frac {2i}3 A^3+ d x^k d x^nd x^m \epsilon_{mnkl}{\rm Tr} \,\lambda\sigma^l\bar\lambda\right)\,.
$$
Finally, the complex three-form
${\mathcal C}_3^0$
  has the following form
  \be\label{C03}
\mathcal C^0_3=\,  {i} E^a \wedge d\theta^\alpha \wedge d\bar\theta^{\dot\alpha}  \sigma_{a\alpha\dot\alpha}\,\theta^2  -\frac 12  E^b\wedge E^a \wedge  d\theta^\alpha
\sigma_{ab\; \alpha}{}^{\beta}{\theta}_{\beta} .
\ee

\subsection{Kappa-symmetry and a supersymmetric static membrane in SYM}

The action \eqref{susym1} is invariant under the fermionic kappa-symmetry (a worldvolume counterpart of supersymmetry) under which the imbedding super-coordinates of the membrane are transformed as follows
\be\label{kappasymm}
\delta\theta^\alpha=\kappa^\alpha(\xi),\qquad \delta \bar\theta^{\dot\alpha}=\bar\kappa^{\dot\alpha}(\xi),\qquad \delta x^m=i\kappa\sigma^m\bar\theta-i\theta\sigma^m\bar\kappa.
\ee
The fermionic  parameters $\kappa_\alpha$ and $\bar \kappa_{\dot\alpha}=(\kappa_\alpha)^*$ are restricted by the following condition
\be\label{kappaproj}
\kappa_\alpha=-{i}\frac {kS+c}{|kS+c|}\Gamma_{\alpha\dot\alpha}\bar\kappa^{\dot\alpha}  \quad \Leftrightarrow \quad \bar\kappa_{\dot\alpha}=-{i}\frac {k\bar S +\bar c}{|kS+c|}\Gamma_{\alpha\dot\alpha}\kappa^{\alpha},
\ee
where
\be\label{kappagamma}
\Gamma_{\alpha\dot\alpha}:= \frac{i\,\epsilon^{ijk}}{3!\sqrt{-\det  h}}\epsilon_{abcd} E^b_iE^c_j E^d_k\,\sigma^a_{\alpha\dot\alpha}, \qquad \Gamma_{\alpha\dot\alpha}\Gamma^{\dot\alpha\beta}=\delta_\alpha^\beta.
\ee
Therefore, kappa-symmetry gauges away 2 of 4 fermionic modes $\theta^\alpha(\xi)$, $\bar\theta^{\dot\alpha}(\xi)$ of the membrane. Since the membrane action is manifestly invariant under worldvolume diffeomorphisms, the latter allow one to gauge fix 3 of 4 bosonic modes $x^{m}(\xi)$. The remaining $3d$ scalar mode $\phi(\xi)$ and a two-component $SL(2, \mathbb R)$ Majorana spinor $\nu_\alpha(\xi)$ form a Goldstone $\mathcal N=1$, $d=3$ supermultiplet associated with partial breaking of $\mathcal N=1$, $d=3$ supersymmetry by the membrane.

It can be shown \cite{Bandos:2019qok} that for a static membrane configuration for which the goldstone fields are equal to zero the membrane action reduces to
that of an $\mathcal N=1$, $d=3$ $SU(N)$ Chern-Simons theory of level $-k$ (in the conventions of \cite{Bashmakov:2018ghn}) induced on the membrane worldvolume by its coupling to the SYM
\be\label{susymstatic}
S_{\rm static}=-\frac{i k}{4\pi}\int d^3\xi{\rm Tr\,}\psi^\alpha\psi_\alpha
+\frac k{4\pi}\int{\rm Tr\,} \left(Ad A+\frac {2i}3 A^3\right)- \frac {|c|}{4\pi}\int \,d^3\xi \,,
\ee
where $A_i(\xi)$ is the induced CS vector field and  $\psi_\alpha(\xi)$ is a 2-component Majorana spinor composed of the real and imaginary component $\lambda_1=\frac 12(\psi_1+i\,\psi_2)$ of the gluino. The integrand in the last term in \eqref{susymstatic} is constant and can be consistently removed (e.g. by sending $c\to 0$ at this stage).

\section{BPS domain wall solutions sourced by the membrane in the VY effective theory}
Let us now consider how the presence of the membrane modifies the equations of motion of the VY theory and induces BPS domain wall solutions \cite{Bandos:2019qok}. We will consider a static membrane whose worldvolume is extended along three space-time directions $\xi^i=x^i$ ($i$=0,1,2) and sitting at $x^3=0=\theta^\alpha=\bar\theta^{\dot\alpha}$. We are interested in solutions for which the fermionic VY field $\chi_\alpha$ vanishes.
Then the membrane action \eqref{susym1}, in which we set $c=0$ reduces to
\be\label{bm}
S_{\rm static}=-\frac 1{4\pi}\int d^3\xi \left(|ks(\xi,0)| {+} kC^3\right),
\ee
where $C^3(\xi,0)= \frac 16 \varepsilon^{ijk}\,C_{ijk}(\xi,0)$.

Adding this action to the bosonic VY action \eqref{bosonicVYL}, we get
\be\label{VY+m}
S=\int d^3\xi dx^3\,\left({\mathcal L}_{\rm VY}^{\rm bos}-\frac 1{4\pi}\delta(x^3) \left(|ks| {+} kC^3\right)\right).
\ee
Varying this action with respect to $\hat D$ and $C^m$ we get modified equations of motions whose solution is (compare with \eqref{Fexpress})
\be\label{Fexpress1}
F\equiv\hat D+{\rm i} \partial_mC^m=-\frac {\partial_{\bar s}({\overline W}+{{\rm i}\,\frac{n+k\Theta(x^3)}{8\pi}}\bar s)}{K_{s\bar s}},
\ee
where $\Theta({x^3})$ is the step function at the point $x^3=0$ at which the membrane sits. We see that the value of $n$ gets shifted by $k$ units when we cross the membrane. This indicates that the membrane of charge $k$ separates two vacua labeled by $n$ and $n+k$, respectively.

The equation of motion of the scalar field $s(\xi)$ has the following form
\be\label{seq}
\Box s\, K_{s\bar s}+ \partial_m s \partial^m s\,  K_{ss\bar s}+ F\bar F  K_{s\bar s\bar s}+\bar F \bar W_{\bar s\bar s}= \frac k{8\pi} \delta (x^3) \frac {ks}{|ks|}\,,
\ee
where it is understood that $F$ is given by \eqref{Fexpress1}.

We look for solutions of \eqref{seq} which describe BPS domain walls that preserve 1/2 supersymmetry and
interpolate between two vacua, which are reached as $x^{3}\to -\infty$ and $x^{3}\to +\infty$ and are separated by the membrane, i.e.
$$
\langle{s}\rangle_{-\infty}=\Lambda^3 e^{\frac{2\pi i n}N }\quad {\rm and }\quad
\langle{s}\rangle_{+\infty}=\Lambda^3 e^{\frac{2\pi i (n+k)}N }\,.
$$
 To this end we follow the well known method (see e.g. \cite{Abraham:1990nz,Dvali:1996xe,Shifman:2009zz}). Namely, the field $s$ is assumed to depend only on the coordinate $x^3$ orthogonal to the membrane and the supersymmetry variation of the fermionic field $\chi_\alpha$ vanishes
\be\label{susyv}
\delta \chi_\alpha = {\tt i}\,\dot s\,\sigma^3_{\alpha\dot\alpha}\bar\epsilon+ \epsilon_\alpha F=0,
\ee
where $\dot s\equiv \partial_3 s\equiv \frac \partial {\partial{x^3}}s$.
On the other hand, for the static membrane we have $\theta^\alpha=\bar\theta^{\dot\alpha}=0$. These conditions are preserved by a combined supersymmetry and kappa-symmetry transformations of $\theta$
\be\label{s+k}
\delta\theta^\alpha=\epsilon^\alpha+\kappa^\alpha=0 \quad \Rightarrow \quad \epsilon^\alpha=-\kappa^\alpha.
\ee
Now remember that the kappa-symmetry parameters are restricted by the condition \eqref{kappaproj} (with $c=0$) which reduces the number of independent real components to two. Therefore, also the supersymmetry parameters should satisfy the very same condition \eqref{kappaproj}. For the static membrane case under consideration, this reduces to
\be
\label{DW_proj}
 \epsilon_\alpha=e^{i \alpha}	\sigma^3_{\alpha \dot \alpha} \bar\epsilon^{\dot \alpha} ,\qquad e^{i \alpha}:=\frac {ks(0)}{|ks(0)|}\,.
\ee
Substituting this relation into \eqref{susyv} we get the BPS equation for the field $s$
\be\label{BPSe}
\dot s= i e^{i\alpha} F %=-i e^{i\alpha} \frac{\overline{\hat W}_{\bar s}}{K_{s\bar s}},
\ee
in which it is understood that $F$ is given by \eqref{Fexpress1}. One can check that this BPS relation solves the field equation  \eqref{seq}.
From \eqref{BPSe} it also follows that
\be\label{dReW=0}
\frac d{d x^3} {\rm Re}\, (\hat We^{-i\alpha})=0\,, \qquad \hat W(s)\equiv W(s)-  \frac i{8\pi} (n+k\Theta(x^3))\,s
\ee
that is
\be\label{ReW=c}
{\rm Re}\, (\hat We^{-i\alpha})={\tt const}
\ee
at each point along $x^3$ including the position of the membrane ($x^3=0$), which is compatible with the worldvolume field equation of $x^3(\xi)$ that for the static membrane has the following form
$$
(\partial_3|ks|+k\partial_mC^m)|_{x^3=0}=0.
$$
Substituting into \eqref{BPSe} the form of  $K$ and $W$ of the Veneziano-Yankielowiz model, eqs. \eqref{K=} and \eqref{VYsp}, we get the explicit form of the BPS equation
\be\label{eBPSe}
\dot s={9}{\rm i}\,\rho N|s|^{\frac 43} e^{{\rm i}\,\alpha}\left({\ln\frac{\Lambda^3}{|s|}+{\rm i}\, \arg s}-\frac{2\pi {\rm i}}N( n+k\Theta(x^3))\right).
\ee
Using this equation we can now compute the BPS value of the on-shell action \eqref{VY+m} of the VY model coupled to the membrane. The result \cite{Bandos:2019qok} gives the correct value \eqref{dwT} of the BPS domain wall, namely
\be\label{onshellS}
S_{\rm BPS} = S_{\rm VY}+S_{\rm membr} =-\,2\, \int d^3 \xi\left|{W}_{x^3\to+\infty}-{W}_{x^3\to-\infty}\right|
\ee
and
\be\label{Tdw}
T_{\rm DW}=T_s+T_{\rm membr}= 2\,\left|{W}_{+\infty}-{W}_{-\infty}\right|=
\frac {N}{8\pi^2}\,\Lambda^3\left|e^{2\pi\,i\,\frac{n+k}N}-e^{2\pi\,i\,\frac{n}N}\right|\,.
\ee
It should be stressed that this value of the tension comprises the contribution of the energy density of the scalar field $s$ and the membrane tension
\be\label{TM:=}
T_M=\frac{|ks(0)|}{4\pi}.
\ee
Without the latter the domain wall tension would not saturate the BPS bound, as the one estimated in \cite{Kogan:1997dt}.

We are now ready to present the explicit form of 1/2 BPS domain walls in the VY description of the $SU(N)$ SYM.

\section{Shape of 1/2 BPS domain walls}
\begin{figure}[h!t]
    \centering
	\includegraphics[width=7.5cm]{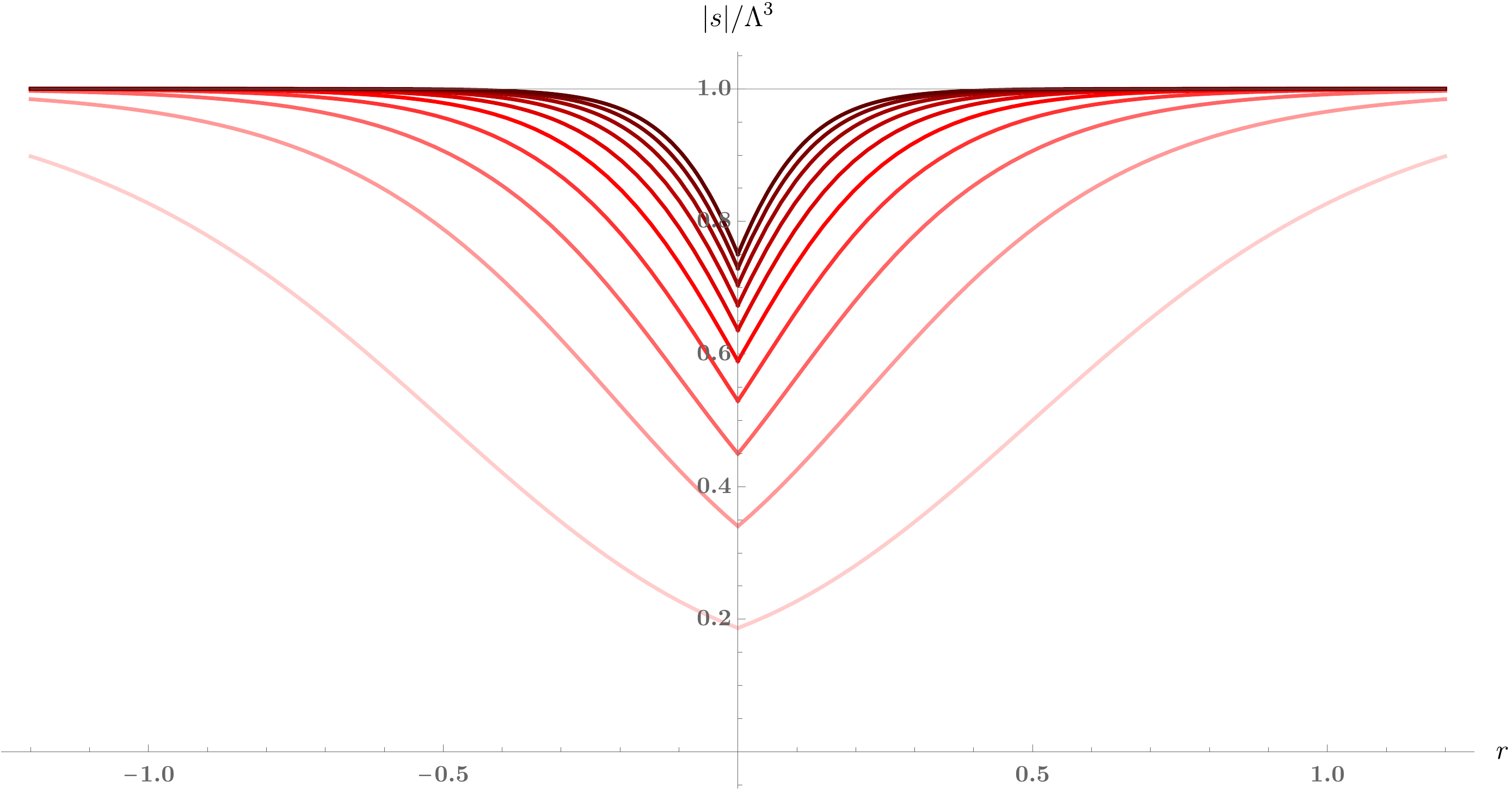} \includegraphics[width=7.5cm]{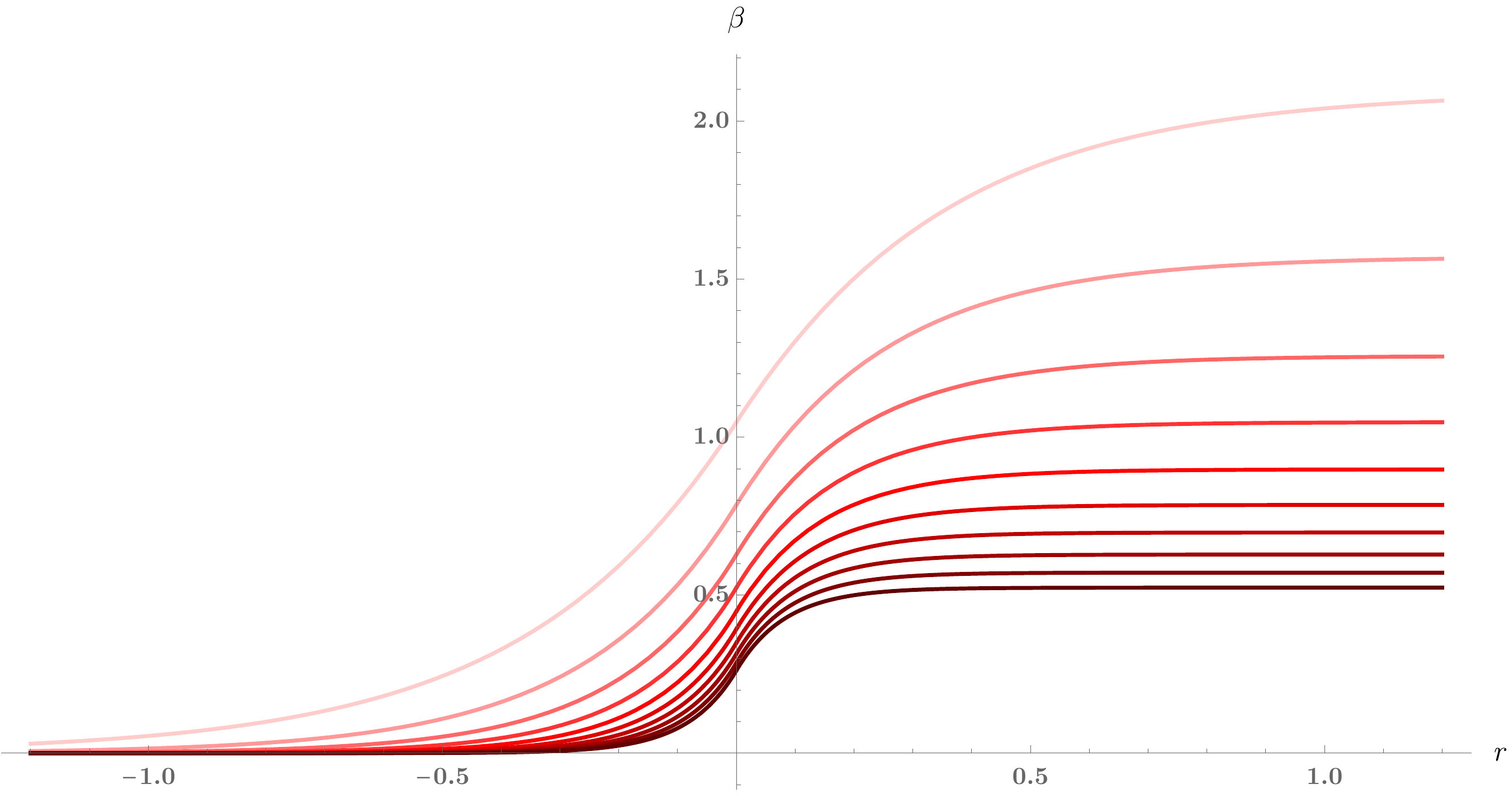}

    \caption{\footnotesize Flow of $\frac{|s|}{\Lambda^3}$ (on the left) and the phase $\beta(x^3)$ (on the right) along $r=9\rho \Lambda x^3$ for different values of $N$, with fixed $k=1$. $N$ are chosen in the interval $[3,12]$, with darker colors corresponding to larger $N$ (alternatively, one might keep $N$ fixed and vary $k$). $\frac{|s|}{\Lambda^3}$ takes the vacuum value $1$ at $x^3=\pm\infty$, decreases and has a cusp at $x^3=0$ where the membrane is sitting.  The flow of $\beta$, starts form $\beta_{-\infty} = 0$ on the left, passes through $\beta(0) = \frac{\pi}{N}$ on the membrane and reaches $\beta_{+\infty} = \frac{2\pi}{N}$ on the right. Thickness of the domain wall solutions decreases when $N$ increases. This can be fixed by choosing $\rho=\frac 1N$ in the K\"ahler potential.}\label{fig:LargeN_flows}
\end{figure}

\begin{figure}[h!t]
    \centering
	\includegraphics[width=9cm]{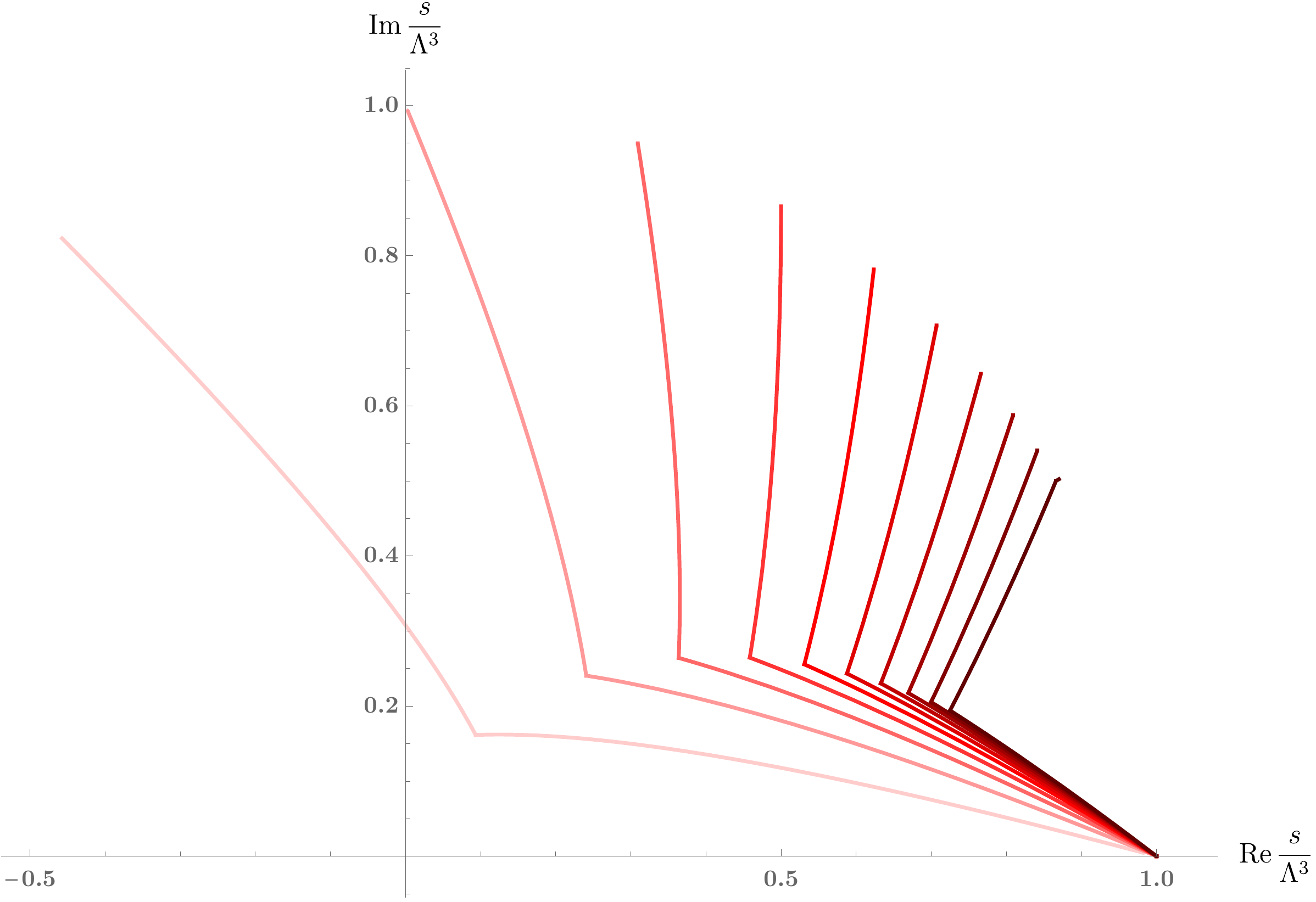}
	 \caption{\footnotesize{Behaviour of $s$ along $x^3 \in(-\infty,+\infty)$ in the complex plane (for $k=1$ and $N$ varying from 3 to 12).  Darker colors correspond to larger $N$. At the point where the membrane is located, $s(x^3)$ has  a cusp.} }\label{fig:LargeN_s}
\end{figure}

\begin{figure}[h!t]
    \centering
	\includegraphics[width=7.5cm]{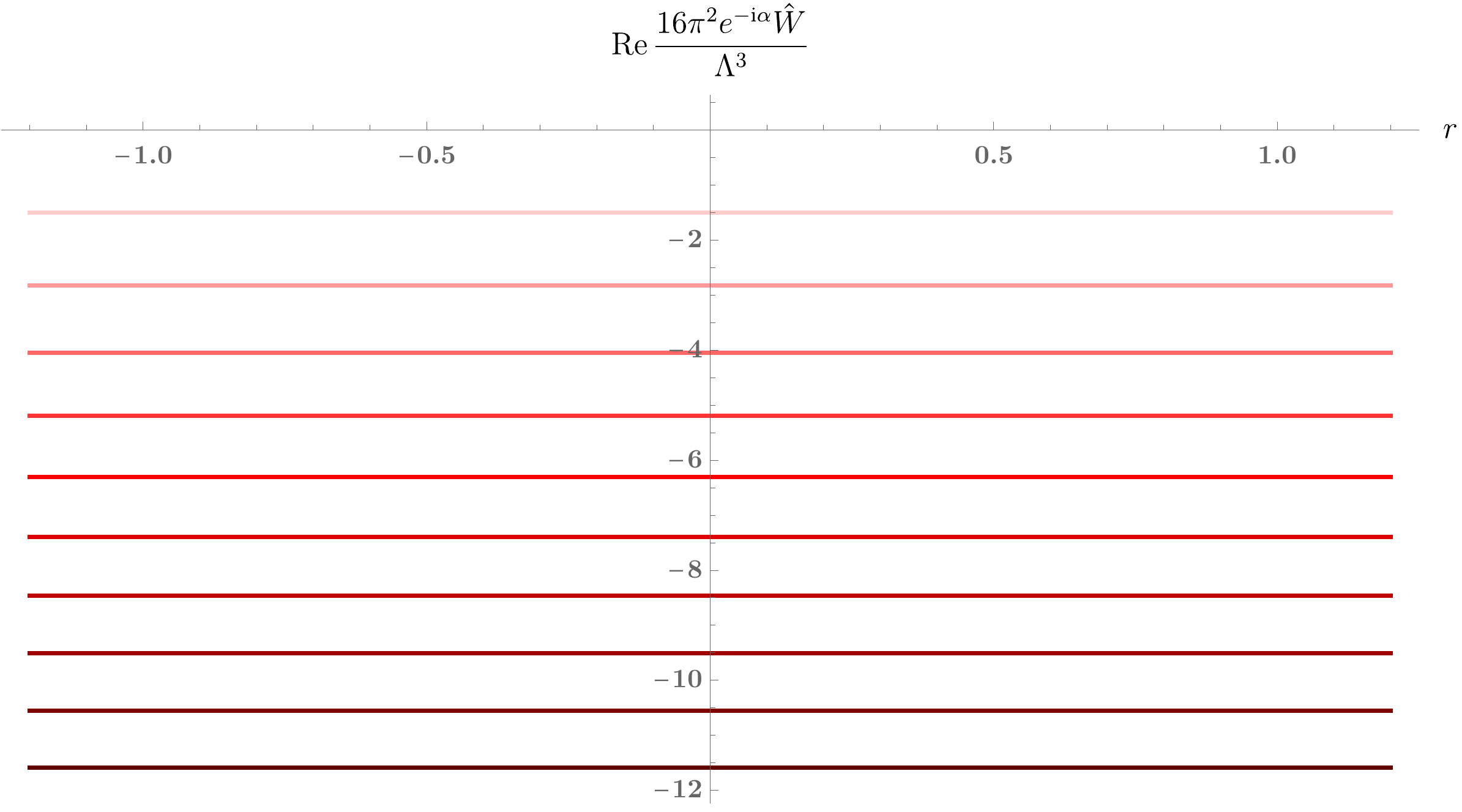} \includegraphics[width=7.5cm]{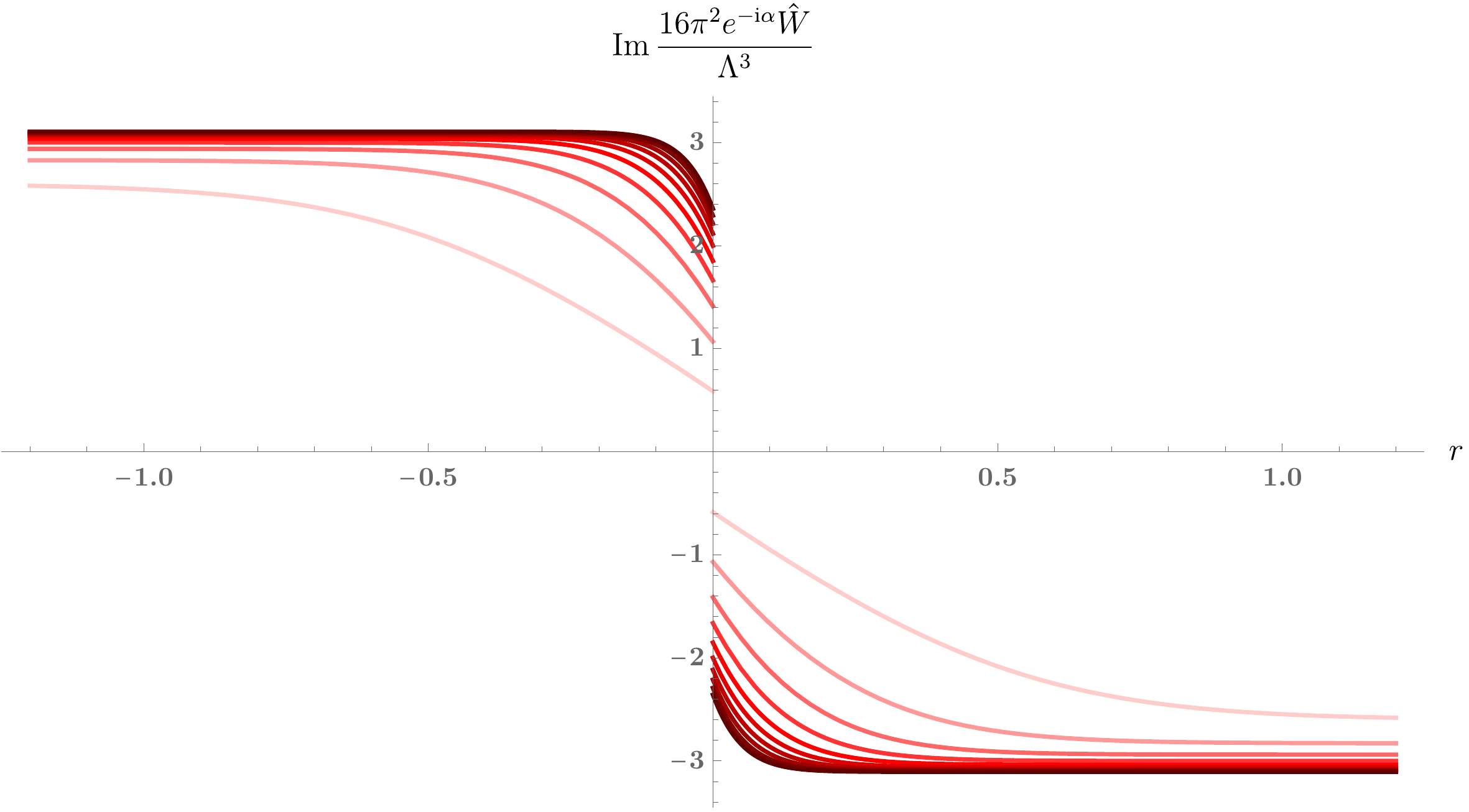}
	 \caption{\footnotesize{Behaviour (for $k=1$ and $N$ varying from 3 to 12) of the real and imaginary part of $\frac{16\pi^2\hat W e^{-i\alpha }}{\Lambda^3}$ ($\alpha=\frac {\pi(1+2n)}N$) along $r=9\rho \Lambda x^3$.  Darker colors correspond to larger $N$. The `jump' of $\Im\frac{16\pi^2\hat W e^{-i\alpha }}{\Lambda^3}$ depicted on the right is proportional to the membrane tension \eqref{TM:=}.} }\label{fig:LargeN_ReW_ImW}
\end{figure}

In \cite{Bandos:2019qok} continuous $s(x)$-field solutions of the BPS equation \eqref{eBPSe} were found for the following values of the membrane charge $k$
$$
|k|\leq \frac N 3.
$$
For $|k|\leq \frac N 3$ the domain walls have the profiles given in Figures  \ref{fig:LargeN_flows}, \ref{fig:LargeN_s} and  \ref{fig:LargeN_ReW_ImW}.
The behaviour of the modulus $|s(x^3)|$ and the phase $\beta(x^3)=\arg s -\frac{2\pi n}{N}$ are given in Figures \ref{fig:LargeN_flows} and \ref{fig:LargeN_s}, and the behaviour of "jumping" superpotential is given in Figure \ref{fig:LargeN_ReW_ImW}. The profiles of these SYM BPS domain walls are similar to those obtained in $\mathcal N=1$ $SU(N)$ super-QCD with $N_f\leq\frac N3$ (where $N_f$ is the number of flavours) in the limit  $m\to\infty$ of the mass of the flavour multiplets \cite{Smilga:2001yz}. From this perspective the membrane may be viewed as an artefact of integrated-out massive flavour modes.

\noindent

As one can see from the plot of $\frac{|s|}{\Lambda^3}$ in Fig.~\ref{fig:LargeN_flows}, it tends to reach zero for smaller $N\geq 3$. As a result the solution breaks down for $N=2$, or equivalently for $k=\frac N2$. This indicates that the domain walls induced by the membranes with a large three-form charge \hbox{$|k|\geq N/2$} should be regarded as  strongly coupled systems, whose internal structure is not captured by the VY effective theory. Note that the cases with $k\leftrightarrow N-k$ are dual to each other since the sum of the charges of the membranes with charge $k$ and $N-k$ is $N$, i.e. equal to the periodicity of the SYM vacua. If $k\leq \frac N3$ then $N-k\geq \frac {2N}3$ and the corresponding dual configurations carry  large three-form charges, and are strongly coupled in this sense.

\if{}
\begin{figure}[h!t]
    \centering
	\includegraphics[width=7.5cm]{Nm_Mod.pdf} \includegraphics[width=7.5cm]{Nm_Phase.pdf}

	\caption{\footnotesize Case $k=-N$.  Flow of $\frac{|s|}\Lambda^3$ (left plot) and $\beta$ (right plot) along the direction $r=9\rho \Lambda x^3$ transverse to the membrane. The solid red  line denotes the field variation on the left of the membrane and the dashed blue line on its right. }
	\label{fig:Nm_flowbetas}
\end{figure}

\begin{figure}[h!t]
    \centering
	\includegraphics[width=9cm]{Nm_S.pdf}
	
	\caption{\footnotesize Case $k=-N$. Variation of the field $s$ with $x^3$ on the complex plane, on the left (red solid line) and on the right (blue dashed line) of the membrane. The field $s$ starts at $s_{-\infty}=1$, reaches the value $s(0) \simeq -3.6$ on the membrane and then flows back to its original value $s_{+\infty}=1$.
	}
	
	\label{fig:Nm_s}
\end{figure}

In the case $|k|=N$ the domain wall interpolates between the two regions separated by the membrane which asymptotically have the same vacuum. So the total tension of this configuration is zero. It would be interesting to understand stability properties of this half-BPS domain wall system.
It is somewhat surprising that for a continuously varying $\beta$ the  solution exists only for $k=-N$  (see \cite{Bandos:2019qok} for more details).  The behaviour of $s(x^3)$ is shown in the Figures \ref{fig:Nm_flowbetas} and \ref{fig:Nm_s}.
The phase $\beta$ continuously varies through the membrane in the interval from 0 to $-2\pi$ and the modulus $|s|$ varies accordingly, as shown in Figure \ref{fig:Nm_flowbetas}. On the complex plane the field $s$ makes a closed loop (Figure \ref{fig:Nm_s}).
\fi

\section{Conclusion}
We have reviewed the construction of the supersymmetric and kappa-invariant action which describes the coupling of a membrane to $\mathcal N=1$, $D=4$ SYM and its Veneziano-Yankielowicz effective generalized sigma-model. We have shown that in the framework of the VY model the presence of the dynamical membrane is required for the formation of 1/2 BPS domain walls interpolating between different SYM gluino condensate vacua and obtained explicit continuous domain wall configurations for $|k|\leq \frac N3$.

More details on the construction and solutions described in this contribution may be found in \cite{Bandos:2019qok}, where it was also shown that a BPS equation similar to \eqref{eBPSe} does not have continuous $s$-field BPS domain wall solutions which might be formed by separated parallel membranes. This indicates that to form a BPS domain wall the membranes should form a stack of coincident membranes, i.e. a composite membrane of a total charge $k$ effectively described by the supermembrane action \eqref{susym1}.  Other problems considered in \cite{Bandos:2019qok} include the construction of a system of an open membrane with a string attached to its boundary. This dynamical brane system was coupled to a massive three-form superfield extension \cite{Burgess:1995kp,Farrar:1997fn} of the Veneziano-Yankielowicz theory, which may be applied to the study of domain-wall junctions.

An interesting issue which requires further study is the relation of the membrane worldvolume action \eqref{susym1} (for the membrane charge $k>1$) to $3d$ gauge theories associated with domain walls in SYM and SQCD (see \cite{Bashmakov:2018ghn} for a review and an exhaustive list of references). As was discussed in \cite{Bandos:2019qok}, for the case $k=1$ our supermembrane action (with the Goldstone fields switched off), eq. \eqref{susymstatic},  is level-rank dual to a corresponding Acharya-Vafa theory \cite{Acharya:2001dz} which was constructed with the use of a stack of $k$ D4-branes wrapped on an internal 2-cycle with $N$ RR fluxes in type IIA string theory.

\acknowledgments
D.S. is grateful to the Organizers of the Corfu Summer Institute 2019 for invitation to present this work and for hospitality during his stay in Corfu. Work of I.B. was supported in part by the Spanish MINECO/FEDER (ERDF EU)  grant  PGC2018-095205-B-I00, by the Basque Government Grant IT-979-16, and the Basque Country University program UFI 11/55. Work of S.L. is funded by a fellowship of Angelo Della Riccia Foundation, Florence and a fellowship of Aldo Gini Foundation, Padova.

\providecommand{\href}[2]{#2}\begingroup\raggedright\endgroup


\begin{thebibliography}{99}

\bibitem{Wess:1974jb}
J.~Wess and B.~Zumino, \emph{{Supergauge Invariant Extension of Quantum
  Electrodynamics}},
  \href{https://doi.org/10.1016/0550-3213(74)90112-6}{\emph{Nucl. Phys.}
  {\bfseries B78} (1974) 1}.

\bibitem{Ferrara:1974pu}
S.~Ferrara and B.~Zumino, \emph{{Supergauge Invariant Yang-Mills Theories}},
  \href{https://doi.org/10.1016/0550-3213(74)90559-8}{\emph{Nucl. Phys.}
  {\bfseries B79} (1974) 413}.

\bibitem{Salam:1974ig}
A.~Salam and J.~A. Strathdee, \emph{{Supersymmetry and Nonabelian Gauges}},
  \href{https://doi.org/10.1016/0370-2693(74)90226-3}{\emph{Phys. Lett.}
  {\bfseries 51B} (1974) 353}.

\bibitem{Witten:1982df}
E.~Witten, \emph{{Constraints on Supersymmetry Breaking}},
  \href{https://doi.org/10.1016/0550-3213(82)90071-2}{\emph{Nucl. Phys.}
  {\bfseries B202} (1982) 253}.

\bibitem{Shifman:1987ia}
M.~A. Shifman and A.~I. Vainshtein, \emph{{On Gluino Condensation in
  Supersymmetric Gauge Theories. SU(N) and O(N) Groups}},
  \href{https://doi.org/10.1016/0550-3213(88)90680-3}{\emph{Nucl. Phys.}
  {\bfseries B296} (1988) 445}.

\bibitem{Davies:1999uw}
N.~M. Davies, T.~J. Hollowood, V.~V. Khoze and M.~P. Mattis, \emph{{Gluino
  condensate and magnetic monopoles in supersymmetric gluodynamics}},
  \href{https://doi.org/10.1016/S0550-3213(99)00434-4}{\emph{Nucl. Phys.}
  {\bfseries B559} (1999) 123}
  [\href{https://arxiv.org/abs/hep-th/9905015}{{\ttfamily hep-th/9905015}}].

\bibitem{Dvali:1996xe}
G.~R. Dvali and M.~A. Shifman, \emph{{Domain walls in strongly coupled
  theories}}, \href{https://doi.org/10.1016/S0370-2693(97)00808-3,
  10.1016/S0370-2693(97)00131-7}{\emph{Phys. Lett.} {\bfseries B396} (1997) 64}
  [\href{https://arxiv.org/abs/hep-th/9612128}{{\ttfamily hep-th/9612128}}].

\bibitem{Bashmakov:2018ghn}
V.~Bashmakov, F.~Benini, S.~Benvenuti and M.~Bertolini, \emph{{Living on the
  walls of super-QCD}},
  \href{https://doi.org/10.21468/SciPostPhys.6.4.044}{\emph{SciPost Phys.}
  {\bfseries 6} (2019) 044} [\href{https://arxiv.org/abs/1812.04645}{{\ttfamily
  1812.04645}}].

\bibitem{Kogan:1997dt}
I.~I. Kogan, A.~Kovner and M.~A. Shifman, \emph{{More on supersymmetric domain
  walls, N counting and glued potentials}},
  \href{https://doi.org/10.1103/PhysRevD.57.5195}{\emph{Phys. Rev.} {\bfseries
  D57} (1998) 5195} [\href{https://arxiv.org/abs/hep-th/9712046}{{\ttfamily
  hep-th/9712046}}].

\bibitem{Bandos:2019qok}
I.~Bandos, S.~Lanza and D.~Sorokin, \emph{{Supermembranes and domain walls in
  $\mathcal N=1$, $D=4$ SYM}},
  \href{https://doi.org/10.1007/JHEP12(2019)021}{\emph{JHEP} {\bfseries 12}
  (2019) 021} [\href{https://arxiv.org/abs/1905.02743}{{\ttfamily
  1905.02743}}].

\bibitem{Veneziano:1982ah}
G.~Veneziano and S.~Yankielowicz, \emph{{An Effective Lagrangian for the Pure
  N=1 Supersymmetric Yang-Mills Theory}},
  \href{https://doi.org/10.1016/0370-2693(82)90828-0}{\emph{Phys. Lett.}
  {\bfseries 113B} (1982) 231}.

\bibitem{Burgess:1995kp}
C.~P. Burgess, J.~P. Derendinger, F.~Quevedo and M.~Quiros, \emph{{Gaugino
  condensates and chiral linear duality: An Effective Lagrangian analysis}},
  \href{https://doi.org/10.1016/0370-2693(95)00183-L}{\emph{Phys. Lett.}
  {\bfseries B348} (1995) 428}
  [\href{https://arxiv.org/abs/hep-th/9501065}{{\ttfamily hep-th/9501065}}].

\bibitem{Binetruy:1996xw}
P.~Binetruy, F.~Pillon, G.~Girardi and R.~Grimm, \emph{{The Three form
  multiplet in supergravity}},
  \href{https://doi.org/10.1016/0550-3213(96)00370-7}{\emph{Nucl. Phys.}
  {\bfseries B477} (1996) 175}
  [\href{https://arxiv.org/abs/hep-th/9603181}{{\ttfamily hep-th/9603181}}].

\bibitem{Gates:1980ay}
S.~J. Gates, Jr., \emph{{Super P-Form Gauge Superfields}},
  \href{https://doi.org/10.1016/0550-3213(81)90225-X}{\emph{Nucl. Phys.}
  {\bfseries B184} (1981) 381}.

\bibitem{Shore:1982kh}
G.~M. Shore, \emph{{Constructing Effective Actions for $N=1$ Supersymmetry
  Theories. 1. Symmetry Principles and Ward Identities}},
  \href{https://doi.org/10.1016/0550-3213(83)90544-8}{\emph{Nucl. Phys.}
  {\bfseries B222} (1983) 446}.

\bibitem{Kovner:1997im}
A.~Kovner and M.~A. Shifman, \emph{{Chirally symmetric phase of supersymmetric
  gluodynamics}}, \href{https://doi.org/10.1103/PhysRevD.56.2396}{\emph{Phys.
  Rev.} {\bfseries D56} (1997) 2396}
  [\href{https://arxiv.org/abs/hep-th/9702174}{{\ttfamily hep-th/9702174}}].

\bibitem{Groh:2012tf}
K.~Groh, J.~Louis and J.~Sommerfeld, \emph{{Duality and Couplings of
  3-Form-Multiplets in N=1 Supersymmetry}},
  \href{https://doi.org/10.1007/JHEP05(2013)001}{\emph{JHEP} {\bfseries 05}
  (2013) 001} [\href{https://arxiv.org/abs/1212.4639}{{\ttfamily 1212.4639}}].

\bibitem{Farakos:2017jme}
F.~Farakos, S.~Lanza, L.~Martucci and D.~Sorokin, \emph{{Three-forms in
  Supergravity and Flux Compactifications}},
  \href{https://doi.org/10.1140/epjc/s10052-017-5185-y}{\emph{Eur. Phys. J.}
  {\bfseries C77} (2017) 602}
  [\href{https://arxiv.org/abs/1706.09422}{{\ttfamily 1706.09422}}].

\bibitem{Lanza:2019nfa}
S.~Lanza, \emph{{Exploring the Landscape of effective field theories}}, Ph.D.
  thesis, Padua U., 2019.
\newblock \href{https://arxiv.org/abs/1912.08935}{{\ttfamily 1912.08935}}.

\bibitem{Bergshoeff:1987cm}
E.~Bergshoeff, E.~Sezgin and P.~K. Townsend, \emph{{Supermembranes and
  Eleven-Dimensional Supergravity}},
  \href{https://doi.org/10.1016/0370-2693(87)91272-X}{\emph{Phys. Lett.}
  {\bfseries B189} (1987) 75}.

\bibitem{Townsend:1987yy}
P.~K. Townsend, \emph{{Supersymmetric extended solitons}},
  \href{https://doi.org/10.1016/0370-2693(88)90852-0}{\emph{Phys. Lett.}
  {\bfseries B202} (1988) 53}.

\bibitem{Abraham:1990nz}
E.~R.~C. Abraham and P.~K. Townsend, \emph{{Intersecting extended objects in
  supersymmetric field theories}},
  \href{https://doi.org/10.1016/0550-3213(91)90093-D}{\emph{Nucl. Phys.}
  {\bfseries B351} (1991) 313}.

\bibitem{Dixon:1991xz}
J.~A. Dixon, M.~J. Duff and E.~Sezgin, \emph{{The Coupling of Yang-Mills to
  extended objects}},
  \href{https://doi.org/10.1016/0370-2693(92)90391-G}{\emph{Phys. Lett.}
  {\bfseries B279} (1992) 265}
  [\href{https://arxiv.org/abs/hep-th/9201019}{{\ttfamily hep-th/9201019}}].

\bibitem{Dixon:1992qd}
J.~A. Dixon and M.~J. Duff, \emph{{Chern-Simons forms, Mickelsson-Faddeev
  algebras and the p-branes}},
  \href{https://doi.org/10.1016/0370-2693(92)90799-A}{\emph{Phys. Lett.}
  {\bfseries B296} (1992) 28}
  [\href{https://arxiv.org/abs/hep-th/9205099}{{\ttfamily hep-th/9205099}}].

\bibitem{Townsend:1993wy}
P.~K. Townsend, \emph{{Effective description of axion defects}},
  \href{https://doi.org/10.1016/0370-2693(93)91499-D}{\emph{Phys. Lett.}
  {\bfseries B309} (1993) 33}
  [\href{https://arxiv.org/abs/hep-th/9303171}{{\ttfamily hep-th/9303171}}].

\bibitem{Bandos:2010yy}
I.~A. Bandos and C.~Meliveo, \emph{{Superfield equations for the interacting
  system of D=4 N=1 supermembrane and scalar multiplet}},
  \href{https://doi.org/10.1016/j.nuclphysb.2011.03.010}{\emph{Nucl. Phys.}
  {\bfseries B849} (2011) 1} [\href{https://arxiv.org/abs/1011.1818}{{\ttfamily
  1011.1818}}].

\bibitem{Cribiori:2020wch}
N.~Cribiori, F.~Farakos and G.~Tringas, \emph{{Three-forms and Fayet-Iliopoulos
  terms in Supergravity: Scanning Planck mass and BPS domain walls}},
  \href{https://arxiv.org/abs/2001.05757}{{\ttfamily 2001.05757}}.

\bibitem{Shifman:2009zz}
M.~Shifman and A.~Yung, \emph{{Supersymmetric solitons}}, Cambridge Monographs
  on Mathematical Physics. Cambridge University Press, 2009,
  \href{https://doi.org/10.1017/CBO9780511575693}{10.1017/CBO9780511575693}.

\bibitem{Smilga:2001yz}
A.~V. Smilga, \emph{{Tenacious domain walls in supersymmetric QCD}},
  \href{https://doi.org/10.1103/PhysRevD.64.125008}{\emph{Phys. Rev.}
  {\bfseries D64} (2001) 125008}
  [\href{https://arxiv.org/abs/hep-th/0104195}{{\ttfamily hep-th/0104195}}].

\bibitem{Farrar:1997fn}
G.~R. Farrar, G.~Gabadadze and M.~Schwetz, \emph{{On the effective action of
  N=1 supersymmetric Yang-Mills theory}},
  \href{https://doi.org/10.1103/PhysRevD.58.015009}{\emph{Phys. Rev.}
  {\bfseries D58} (1998) 015009}
  [\href{https://arxiv.org/abs/hep-th/9711166}{{\ttfamily hep-th/9711166}}].

\bibitem{Acharya:2001dz}
B.~S. Acharya and C.~Vafa, \emph{{On domain walls of N=1 supersymmetric
  Yang-Mills in four-dimensions}},
  \href{https://arxiv.org/abs/hep-th/0103011}{{\ttfamily hep-th/0103011}}.

\end{thebibliography}
\end{document}